\documentstyle{article}

\newcommand{\ft}[2]{{\textstyle\frac{#1}{#2}}}                                   
\newcommand{\eqn}[1]{(\ref{#1})}

\def\1ad{\mbox{\normalsize $^1$}}
\def\2ad{\mbox{\normalsize $^2$}}
\def\3ad{\mbox{\normalsize $^3$}}
\def\4ad{\mbox{\normalsize $^4$}}
\def\5ad{\mbox{\normalsize $^5$}}
\def\6ad{\mbox{\normalsize $^6$}}
\def\7ad{\mbox{\normalsize $^7$}}
\def\8ad{\mbox{\normalsize $^8$}}
\def\makefront{\vspace*{1cm}\begin{center}
\def\newtitleline{\\ \vskip 5pt}
{\Large\bf\titleline}\\
\vskip 1truecm
{\large\bf\authors}\\
\vskip 5truemm
\addresses
\end{center}
\vskip 1truecm
{\bf Abstract:}
\abstracttext
\vskip 1truecm}
\begin {document}
\begin{center}
\hfill THU-98/01\\
\hfill SWAT-98/181\\[2mm]
\hfill {\tt hep-th/9801039}
\end{center}

\def\titleline{
Special Geometry and Compactification on a Circle  
}
\def\authors{
B. de Wit\1ad, B. Kleijn\1ad  and S. Vandoren\2ad
}
\def\addresses{
\1ad
Institute for Theoretical Physics, Utrecht University, 
3508 TA Utrecht, Netherlands\\
\2ad
Department of Physics, University of Wales Swansea, 
SA2 8PP Swansea, U.K.
}
\def\abstracttext{
We discuss some consequences of our previous work on rigid 
special geometry in hypermultiplets in 4-dimensional Minkowski 
spacetime for supersymmetric gauge dynamics when one of the 
spatial dimensions is compactified  on a circle. 
}
\makefront
\section{Introduction}
Special geometry appears in the context of $N=2$ supersymmetric 
gauge theories in 4 spacetime dimensions. Such theories arise as 
low-energy effective actions of type-II strings compactified
on a Calabi-Yau manifold ${\cal X}$. The vector 
multiplets exhibit special geometry, with the corresponding scalars 
parametrising  a so-called special K\"ahler manifold.
The hypermultiplet moduli space (which is a quaternionic 
space) of type  IIB/A is related to the vector-multiplet special 
K\"ahler manifold of type IIA/B by string duality (at least 
in string perturbation theory). This  
connection was first studied in \cite{CFG}
in terms of the ${\bf c}$-map which converts special K\"ahler into 
quaternionic manifolds. The ${\bf c}$-map can be induced by 
dimensional reduction from 4 to 3 spacetime dimensions.
It relates the classical moduli spaces of vector 
multiplets and hypermultiplets. 
To go beyond this description, one may consider type-II 
strings compactified on ${\cal X}\times  S^1$ and  
perform a T-duality  on the circle \cite{SS}. Such a 
T-duality relates the IIA and IIB strings \cite{ABdual}. When studying the 
4-dimensional theories with a compactified coordinate, it is 
therefore of interest to go beyond a trivial  
dimensional reduction and retain all the 
modes associated with the $S^1$-compactification. 

Motivated by this, we study supersymmetric gauge theories 
with one coordinate compactified on $S^1$. We restrict 
ourselves to rigid $N=2$ supersymmetry throughout. 
In \cite{DJDWKV} we considered the zero-modes of the vector 
multiplets in an $S^1$-compactification for general abelian supersymmetric 
actions. After dualising vector 
multiplets into hypermultiplets, one can thus define a notion 
of (rigid) special geometry in hypermultiplets.  
The aim there was to identify and study the special-geometry 
features for the corresponding hyper-K\"ahler geometries. In 
\cite{DJDWKV} the dynamical effects associated with the 
$S^1$-compactification did not play a role. In this note we 
discuss the 
results of \cite{DJDWKV}, also taking into account certain 
effects caused by the massive modes. It is known that the 
supersymmetric gauge theories exhibit interesting dynamics when 
compactified on $S^1$. In \cite{3dSW} this was analysed for the 
gauge group SU(2). 

Before turning to the hyper-K\"ahler manifolds let us devote the 
remainder of this section to a discussion of vector multiplets 
and symplectic reparametrisations. 
Vector multiplets contain bosonic fields $X^I$ and $A_\mu^I$, with 
$I=1,...,n$ labeling the vector multiplets.
The complex scalar fields $X^I$
parametrise a K\"ahler manifold characterised by a holomorphic 
function $F(X)$. The K\"ahler 
potential is given by
\begin{equation}
K(X,\bar X)=-i{\bar X}^IF_I + i{\bar F}_I X^I\ ,
\end{equation}
where $F_I$ denotes the derivative of $F$ with respect to $X^I$. 
The symplectic group Sp$(2n,{\bf R})$ which acts on the 
(anti-)selfdual components of the field strenghts also 
acts on the scalar fields and on the function $F(X)$. This can be 
expressed by introducing `sections' 
$(X^I,F_I)$ which transform under the symplectic group by 
multiplication with an Sp$(2n,{\bf R})$ matrix,
\begin{equation}
\pmatrix{X^{I}\cr  F_{I}\cr} \longrightarrow  \pmatrix{\tilde 
X^I\cr\tilde F_I\cr}=
\pmatrix{U&Z\cr W&V\cr} \pmatrix{X^{I}\cr  F_I\cr}\,. 
\label{transX}
\end{equation}
{From} this, one can find the transformation of $F_{IJ}$, which 
is of the same form as the transformation of a 
period matrix of a genus-$n$ Riemann surface under a redefinition 
of the homology basis (eventually the 
symplectic reparametrisations will also be restricted to Sp($n,
{\bf Z})$).  
Denoting ${\cal S}^I{}_{\!J}=\partial {\tilde X}^I/\partial X^J$, one has
\begin{equation}
\tilde{F}_{IJ} = (V_I{}^K {F}_{KL}+ W_{IL} )\big[{\cal S}^{-1}]^L{}_J\ .
\end{equation}
{From} these quantities one can construct symplectically 
covariant objects like, for instance, the K\"ahler
potential. Other well-known examples are \cite{DW}
\begin{equation}
C_I=F_I-F_{IJ}X^J=\partial_IG\,, \qquad G=2F-X^IF_I\, ,
\qquad F_{IJK} \,.
\end{equation}
Symplectically covariant means that ${\tilde C}_I(\tilde X) 
=C_J(X) \big[{\cal S}^{-1}]^J{\!}_I$; likewise $F_{IJK}$ 
transforms as a 3-rank tensor and $G$ transforms as a 
function, i.e. $\tilde G(\tilde X) = G(X)$. For a single vector 
multiplet, using standard notation $C_I$ equals  
\begin{equation}
C=a_D-\tau a\ .
\end{equation}
where $\tau$ denotes the second derivative of $F(a)$. 
The matrix $\cal S$ is then equal to ${\cal S}= U + Z \tau$, so 
that, when expressing the functions $C$, $G$  and the third derivative of $F$ 
(which can be written as ${\rm d}\tau/{\rm 
d}a=-1/({\rm d}^2C/{\rm d}\tau^2)$) in terms of $\tau$, the symplectic 
reparametrisations  take the form of modular transformations 
acting on modular forms of weights $-1$, 0 and $-3$, respectively. 
When the U(1) theory describes the Wilsonian effective action of an 
underlying pure SU(2) gauge theory \cite{SW}, the function $G$ is 
{\it invariant} under the subgroup $\Gamma(2)$ of Sp$(2,{\bf Z})$ 
that is associated with the monodromies \cite{DW}.
The function $G$ was studied in \cite{Matone,STY-EY} and shown to 
measure the scaling violation of the underlying microscopic 
theory. The modular form $C$  is the central object in Nahm's approach 
\cite{Nahm} to construct the Seiberg-Witten solution. Knowing the poles
of $C$ as a function of $\tau$ determines it up to a  
multiplicative factor. The periods then follow from
the relation $a=-{\rm d}C/{\rm d}\tau$. A generalisation of this 
approach for  
higher-rank gauge groups has not been given so far. This would 
require the study of automorphic forms, of which much less
is known. These automorphic forms also appear in the context of
$N=2, d=4$ heterotic or type-II string compactifications, see 
e.g. \cite{dWKLL}. 

As was shown in \cite{MMM} (see also \cite{Mat}) $F_{IJK}$ satisfies a 
WDVV-like  \cite{WDVV} equation. Denoting matrices
$(F_I)_{MN}=F_{IMN}$, the equation takes the form 
\begin{equation}
F_IF_K^{-1}F_J=F_JF_K^{-1}F_I\ .
\end{equation}
It is obviously consistent with symplectic transformations 
because $F_{IJK}$ transforms as a tensor.
In the case of SU(2) and SU(3), the equation is trivial.

Because the supercharges are symplectic invariants, central 
charges in the sypersymmetry algebra are also symplectically 
invariant objects. An example of this is the well-known BPS mass, 
which is defined in terms of the electric and magnetic charges 
that transform under symplectic reparametrisations as a 
$2n$-component vector. These charges 
parametrise a certain lattice and the symplectic group must be 
restricted to the integer-valued subgroup that leaves this lattice 
invariant. We return to the central charges in the next  
section.

\section{Special geometry and hyper-K\"ahler manifolds}
In this section, we discuss how symplectic transformations in the 
hyper-K\"ahler manifolds
are induced by those on the vector-multiplet side. As we will 
see, a number of new symplectically covariant 
objects can be defined in terms of hypermultiplets.
Let us first consider the reduction to 3 dimensions, where we 
only retain the zero-modes on $S^1$. We start from a 
4-dimensional (effective) theory of abelian vector multiplets 
based on a function $F(X)$ and a corresponding K\"ahler potential $K$.
Upon dimensional reduction, one obtains new complex scalars $Y_I$ 
originating from the gauge fields, 
\begin{equation}
Y_I=B_I-F_{IJ}A^J\ .
\end{equation}
Here $A^I$ are the gauge fields $A_\mu^I$ with the index $\mu$ 
corresponding to the compactified coordinate.
The scalar fields $B_I$ come from dualising the 3-dimensional 
abelian gauge fields. They are the Lagrange multipliers that 
were introduced to enforce the Bianchi identities.  
Under symplectic transformations $(A^I,B_I)$ transforms as a 
$2n$-component vector (cf. (\ref{transX})). Consequently $Y_I$ 
transforms as a co-vector, 
i.e. ${\tilde Y}_I=Y_J\big[{\cal S}^{-1}]^J{}_I$.
The existence of $N=4$ supersymmetry in 3 dimensions implies 
that the manifold parametrised by the complex scalars $(X^I,Y_I)$ 
must be hyper-K\"ahler. It was shown in \cite{CFG,DJDWKV} that a 
corresponding K\"ahler potential is given by  
\begin{equation}
K(X,Y,\bar X,\bar Y) = -i{\bar X}^IF_I+ i{\bar F}_I 
X^I-\ft12(Y-{\bar Y})_IN^{IJ}(Y-{\bar Y})_J\  ,\label{Kpot}
\end{equation}
which is invariant under symplectic transformations up to K\"ahler 
transformations. Here $N^{IJ}$ denotes the inverse of $N_{IJ}= 
-iF_{IJ} +i\bar F_{IJ}$. 

Assuming that the radius of the circle is equal to $R$, one imposes 
periodic boundary conditions and expands the fields in Fourier 
modes. The massless zero-mode discussed above is the only one 
that survives the (naive)  
$R\rightarrow 0$ limit, as the other modes have masses proportional 
to $1/R$. When studying the effective actions of an underlying 
microscopic theory, these massive modes are integrated out
in the Wilsonian sense. For the moment we will neglect them.
{From} gauge transformations with non-trivial winding
in the compactified direction, it follows that the (zero-mode) 
fields $A^I$  are defined modulo $1/R$; from the periodicity 
of the generalised theta angles it follows that also the fields 
$B_I$ are periodic with the same period (provided one chooses a 
suitable normalisation of the Lagrange multipliers). The fields 
$A^I$ and $B_I$ span a torus $T^{2n}$ above each point of the  
special K\"ahler moduli space whose volume is equal to 
$(4R)^{-n}$. At this point one may identify the torus at a 
given point $X$ in the special K\"ahler space with the Jacobian 
variety of an auxiliary Riemann surface ${\cal M}_X$ that can be 
associated with some underlying 4-dimensional dynamics of a 
nonabelian gauge theory in the Coulomb phase \cite{SW}, with 
its period matrix given by $F_{IJ}(X)$. The 
scalars $Y_I$ take their values in this Jacobian. We now note the 
existence of the following holomorphic one-forms, which are 
manifestly covariant under symplectic reparametrisations,
\begin{equation}
{\cal W}_I= {\rm d} B_I- F_{IJ}\, {\rm d}A^J\,.
\end{equation}
These forms appear in the symplectically covariant 
Sp(1)$\times {\rm Sp}(n)$ one-forms that characterise the 
hypermultiplet couplings and supersymmetry transformations 
\cite{DJDWKV}. A symplectically invariant holomorphic two-form is   
\begin{equation}
\omega={R}\,{\rm d}X^I \wedge {\cal W}_I= 
{R}\,{\rm d}X^I \wedge {\rm d}Y_I 
={R}\Big({\rm d}X^I \wedge {\rm d}B_I - {\rm 
d}F_I\wedge {\rm d}A^I\Big) \ .\label{hol2form}
\end{equation}
This two-form, its conjugate and the K\"ahler
two-form corresponding to (\ref{Kpot}) are symplectically 
invariant and closed. We return to these three hyper-K\"ahler 
forms later.

Assuming that the sections $(X^I,F_I)$ can be 
written in terms of a number of modular parameters $u^\alpha$, we 
consider the following symplectically invariant one-forms on 
the Jacobian variety,
\begin{equation}
\lambda_\alpha  = {R} \,{\partial X^I\over \partial 
u^\alpha} {\cal W}_I = {R} \left({\partial X^I\over 
\partial u^\alpha} {\rm d}B_I - {\partial F_I\over \partial 
u^\alpha} {\rm d}A^I\right) \,.
\end{equation} 
The integrals of these one-forms along the one-cycles $\alpha^I$ 
and $\beta_I$ associated with the coordinates $A^I$ 
and $B_I$,  yield 
\begin{equation} 
\partial_\alpha X^I(u) = \oint_{\beta_I} \lambda_\alpha \,,   \qquad
\partial_\alpha F_I(u) = - \oint_{\alpha^I} \lambda_\alpha \,.   
\label{Jacperiods}
\end{equation} 
The symplectic transformations on the left-hand side of these 
equations are now induced by the transformations of the 
homology cycles that leave the canonical intersection matrix 
invariant. 

The result (\ref{Jacperiods}) can also be formulated in terms of the 
corresponding homology cycles of the underlying Riemann surface. 
In the special case of SU(2), the Jacobian variety and the 
Riemann surface can be identified with the same torus with 
modular parameter $\tau$ \cite{3dSW}. The complex coordinate we 
use on the torus is $Y=B-\tau A$. 
One can easily compute the period ``matrix" from the lengths of
the $A$ and $B$ cycles (see \cite{DJDWKV}). It is independent of 
the compactification radius, 
\begin{equation}
\frac{l_A}{l_B}=|\tau|\ .
\end{equation}
One thus expects that there is 
a relation between the Seiberg-Witten holomorphic one-form 
$\lambda$ and ${\cal W}= {\rm d}B-\tau \,{\rm d}A$.
Following the discussion of \cite{3dSW} (section 3.1), adapted to our 
notation, one has
\begin{equation}
\lambda= \frac{{\rm d}x}{y}={R} \,\frac{{\rm d}a}{{\rm 
d}u}\,{\cal W}= {R}\left(\frac{{\rm d}a}{{\rm d}u}{\rm d}B-
\frac{{\rm d}a_D}{{\rm d}u}{\rm d}A\right)\ ,
\end{equation}
where the torus is parametrised by e.g. $y^2=(x-1)(x+1)(x-u)$. 
The holomorphic two-form $\omega$ 
over the hyper-K\"ahler manifold, defined in (\ref{hol2form}), 
equals 
\begin{equation}
\omega={R}\, {\rm d}a\wedge {\rm d}Y=\frac{{\rm d}u\wedge 
{\rm d}x}{y}\  .
\end{equation}

The hyper-K\"ahler two-forms corresponding to $\omega$, its 
complex conjugate and the K\"ahler form, play an important role in 
the discussion of central charges.
{From} the anticommutator of the supercharges, it follows that both vector 
multiplets and hypermultiplets may be subject to three central charges 
in the susy algebra.
For vector multiplets one has the K\"ahler 
two-form central charge and the (anti-)holomorphic BPS mass.
All three are symplectically invariant.
For hypermultiplets, the central charges are integrals over the 
three hyper-K\"ahler two-forms. 
As was discussed in \cite{DJDWKV}, the central charges in 3 
dimensions are enumerated by the second homotopy group of the 
target-space manifold. For hyper-K\"ahler spaces that are in the image
of the ${\bf c}$-map, they are symplectically
invariant. To realise these central charges explicitly, we take the 
hyper-K\"ahler manifold with the torus $T^{2n}$ fibered over the 
special K\"ahler space. 
The two-forms we have to integrate over are closed and thus locally exact.
Indeed, for the holomorphic two-form $\omega$ one has 
\begin{equation}
{\rm d}X^I\wedge {\rm d}Y_I={\rm d}\left( X^I{\rm d}B_I-F_I{\rm d}A^I\right)\ .
\end{equation}
The corresponding central charge can be written as
\begin{equation}
Z= {R} \int _{C_1} \left(X^I{\rm d}B_I-F_I{\rm d}A^I\right)\ ,
\end{equation}
with $C_1$ a non-vanishing one-cycle in the hyper-K\"ahler space.
A relevant cycle can be chosen on the hyper-torus $T^{2n}$ and 
decomposed in terms of a canonical homology basis of one-cycles 
$\alpha^I$ and $\beta_I$ as 
\begin{equation}
C_1=q_{{\rm e}I}\alpha^I+q_{\rm m}^I\beta_{I}\ ,
\end{equation} 
with integer coefficients $q_{\rm e}$ and $q_{\rm m}$. This leads 
to  
\begin{equation}
Z=X^I\,q_{{\rm e}I}-F_I\,q_{\rm m}^I\ .
\end{equation}
Of course, because of the ${\bf c}$-map, this is precisely the 
central charge of the vector multiplet model
we started with. It illustrates how the torus is related to the lattice of 
electric and magnetic charges.

\section{The effective actions and Kaluza-Klein modes}
We now consider some dynamical effects when compactifying 
the theory on a circle of fixed radius $R$.
For simplicity, we consider the case of pure SU(2) supersymmetric 
Yang-Mills theory. 
In the decompactification limit $R\rightarrow \infty$, the classical plus 
perturbative contributions to the function $F$ that encodes the 
Wilsonian effective action, is given by  
\begin{equation}
F(X)={\textstyle\frac{1}{2}}\tau_0 
X^2+\frac{i}{2\pi}X^2\ln\frac{X^2}{\Lambda^2}\ .\label{prep1loop}
\end{equation}
Here, $\tau_0$ describes the bare coupling and theta parameter 
according to $\tau_0=\theta_0/2\pi+4\pi i/g_0^2$. 
The K\"ahler potential and metric are equal to 
\begin{equation}
K_{4d}=\frac{8\pi}{g_0^2}\, X{\bar X}+\frac{2X{\bar 
X}}{\pi}\bigg[\ln \frac{X{\bar  
X}}{\Lambda ^2}+1\bigg]\,,\qquad
N_{X{\bar X}}=\frac{8\pi}{g_0^2}+\frac{2}{\pi}\left[\ln 
\frac{X{\bar X}}{\Lambda^2}+3\right]\ .
\label{1loop}
\end{equation}
The constant in front of the perturbative corrections is equal to 
the one-loop beta function of SU(2) supersymmetric Yang-Mills 
theory. 

In 4 dimensions, the effective action with at most two 
derivatives is encoded in a 
holomorphic function $F(X)$. This follows from requiring that the 
action depends only on gauge-covariant objects, such as field 
strengths and covariant derivatives. When one of the dimensions 
is compactified on $S^1$, the corresponding zero modes $A^I$ 
associated with the phase of the Wilson line around $S^1$, is 
gauge invariant and can appear in the effective action in a less 
restricted way. The only implication from gauge invariance is 
that the effective action is invariant under shifts of $A^I$ with 
a multiple of $1/R$. The presence of the fields $A^I$ causes a 
change in the holomorphic structure of the various quantities.

To illustrate this more concretely, let us compute the renormalisation 
of the coupling constant in the $S^1$-compactification for SU(2). We first 
note that the unrenormalised coupling constants for the 4- and 
3-dimensional theories, denoted by $g_0$ and $e_0$, 
respectively, are related by 
\begin{equation}
g_0^2=2 \pi R\,e_0^2\ .
\end{equation}
The one-loop contribution to the coupling constant $e$ at finite 
$R$ is given by \cite{dWGR,DKMTV}
\begin{equation}
\frac{1}{e^2}=\frac{1}{e_0^2}+4i \sum _{n=-\infty}^{+\infty}\int\, 
\frac{{\rm d}^3p}{(2\pi)^3} 
\,\Big[p^2-M^2-\Big(A+\frac{n}{R}\Big)^2\Big]^{-2} 
=\frac{1}{e_0^2}-\frac{1}{2\pi} \sum_{n=-\infty}^{+\infty} 
\,\Big[M^2+\Big(A+\frac{n}{R}\Big)^2\Big]^{-\ft12}\ ,  
\label{KKsum}
\end{equation}
where we assume a constant background of the scalar fields, 
$X$ and $A$. Here $M^2=2X{\bar X}$ denotes the mass of the charged 
particles in the 4-dimensional theory. Note the manifest 
periodicity of $A$ in units of $1/R$. In the decompactification 
limit, the above expression yields 
\begin{equation}
\frac{1}{e^2}=\frac{1}{e_0^2}+8i\pi \,R \int\, 
\frac{{\rm d}^4p}{(2\pi)^4}\;
{1\over[p^2-M^2]^2}  = \frac{1}{e_0^2}+{R \over 
2\pi}\log{X\bar X\over \Lambda^2}\,, \label{4dcharge}
\end{equation}
where we have cut off the momentum integral at a scale $\Lambda$, 
chosen such as to coincide with the cut-off in (\ref{1loop}). 
Observe that the $A$-dependence has disappeared, which can be 
understood from the fact that in the uncompactified case, a 
constant vector potential is gauge equivalent to zero. 

For arbitrary value of $R$ we can further evaluate (\ref{KKsum}) by means 
of a Poisson resummation \cite{OV},
\begin{equation} 
\frac{1}{e^2}=\frac{1}{e_0^2}+ {R\over 2\pi} \log{X\bar X\over 
\Lambda^2} - {2 R\over \pi} \,
\sum_{n=1}^{+\infty} K_0(2\pi RM\,n) \cos (2\pi RA\,n)\,,
\end{equation}
where we have again adjusted the cut-off $\Lambda$ such as to 
make contact with (\ref{1loop}). For large $R$, the modified 
Bessel function in the infinite sum vanishes exponentially, so 
that we indeed recover the result (\ref{4dcharge}). 
Note that $2\pi RA$ equals the flux through the $\alpha$-cycle 
of the torus.  

In the limit $R\rightarrow 0$, the $S^1$-modes become infinitely 
heavy and decouple (up to some renormalisation effect), except 
for the $n=0$  
mode. One is left with the 3-dimensional result and \eqn{KKsum} 
yields a charge renormalisation given by
$1/e^2=1/e_0^2-1/(2\pi\sqrt{2X\bar X + A^2})$ \cite{DKMTV}. Here 
we did absorb an infinite renormalisation into the definition of $e_0$. 

The above results demonstrate that the holomorphic structure 
that is characteristic for special geometry, is lost. The equivalence 
transformations of the 4-dimensional theories, which take the 
form of symplectic reparametrizations, are therefore no longer 
manifest in the $S^1$-compactification. An intriguing question 
is, whether these equivalence tranformations can still remain in 
some modified form, for instance, after combining with T-duality. 
This question deserves further study.

\vspace{4mm}

\noindent
{\bf Acknowledgement} We thank J. De Jaegher for her 
collaboration in part of the work reported here. S.V. thanks S. 
Howes and D. Olive for a discussion on ref. \cite{Nahm}.


\end{document}